\documentclass[10pt]{article}

\usepackage[final]{epsfig}
\usepackage{amsmath}
\usepackage{parskip}

\def\lsim{\mathrel{\rlap{\lower4pt\hbox{\hskip1pt$\sim$}}
    \raise1pt\hbox{$<$}}}         
\def\gsim{\mathrel{\rlap{\lower4pt\hbox{\hskip1pt$\sim$}}
    \raise1pt\hbox{$>$}}}         

\def\overleftrightarrow#1{\vbox{\ialign{##\crcr
    $\leftrightarrow$\crcr
    \noalign{\kern 1pt\nointerlineskip}
    $\hfil\displaystyle{#1}\hfil$\crcr}}}

\begin{document}

\hspace{12cm}{\bf TUM/T39-02-18}\\
\hspace*{12cm}{\bf ECT$^\ast$-02-25}\\

\begin{center}
{\bf Statistical Hadronization as a Snapshot of a dynamical Fireball Evolution
\footnote{work supported in part by BMBF, GSI and by the European Commission under
contract HPMT-CT-2001-00370}}
\end{center}
\begin{center}
{Thorsten Renk$^{a}$}

{\small \em $^{a}$ Physik Department, Technische Universit\"{a}t M\"{u}nchen,
D-85747 Garching, GERMANY\\
and ECT$^\ast$, I-38050 Villazzano (Trento), ITALY}

\end{center}
\vspace{0.25 in}

\begin{abstract}
Statistical hadronization models are extremely successful in describing
measured ratios of hadrons produced in heavy-ion collisions
for a wide range of beam energies
from SIS to RHIC.
Using the idea of statistical hadronization at the phase boundary
within the framework of a recently proposed dynamical model for the thermodynamics
of fireballs, we establish a relation between the
equation of state (EoS) in the partonic phase
probed at the phase transition temperature $T_C$ and the measured
hadron ratios. In this way, the ratios can be predicted parameter-free. 
We demonstrate that this framework gives a
consistent description of conditions both at SPS and RHIC.
As the data on dilepton emission from heavy ion collisions 
give evidence for strong in-medium modifications of hadronic
properties, we schematically investigate the sensitivity of our
results to such modifications and qualitatively
sketch a consistent scenario involving such in-medium effects.
\end{abstract}

\vspace {0.25 in}

\section{Introduction}
\label{sec_introduction}

The goal of current investigations in heavy-ion collisions
is to establish the existence of the quark-gluon plasma (QGP)
phase and to study its properties. Several signals have been proposed 
as indicators for
its creation. However, no unambiguous proof has been established so far.
A large part of the problem lies in the difficulty to separate
effects of the evolution of the hot and dense medium from
characteristic changes in the physics of reactions
taking place inside this medium. Therefore, it is mandatory
to aim at a consistent description of as many observables
as possible within a single model of the medium evolution.

In a recent paper \cite{MyDileptons}, we have proposed a model
for the evolution of a fireball assuming local thermal equilibrium
and isentropic expansion. In this approach, the evolution is constrained
by two major pieces of information: first, by 
measured hadronic momentum spectra and Hanbury-Brown Twiss (HBT) interferometry
data which reflect the freeze-out state; secondly, by information on the
Equation of State (EoS) obtained in lattice 
simulations and represented in terms of a 
quasiparticle picture. We have demonstrated that this evolution
scenario is in agreement with data on dilepton emission
measured by the CERES collaboration \cite{CERES}.
We have also used the same model to describe the emission of hard thermal
photons \cite{Photons} and found good agreement with the data measured by
WA98  \cite{PhotonData}.

No explicit statement about the hadrochemical composition
of the fireball is made in this model. Instead, we use phenomenological
arguments to include the effects of enhanced pion phase space
density into the EoS of hot hadronic matter. On the other
hand, statistical models are extremely successful
in describing the measured ratios of different hadron
species for a range of collision energies from SIS to RHIC
(see e.g. \cite{PBM1,PBM2, PBM3, Broniowski}).
It is the aim of this paper to show that our fireball
evolution model is consistent with the idea of statistical
hadronization.

The paper is organized as follows: First, we outline the version
of statistical hadronization used in our model. After
comments on technical details, we present results and compare to
data both for SPS and RHIC conditions.
We then investigate the sensitivity of these results
to the model parameters, specifically we study the role
of possible in-medium modifications of particle properties.
We conclude by sketching a possible scenario in which
in-medium modifiactions are possible while statistical
hadronization is still in agreement with the data.

\section{The statistical hadronization model}

The fireball evolution model outlined in \cite{MyDileptons}
contains a phase transition from partonic to hadronic
matter. If one is interested in the composition of
the hadronic system, one has to introduce
a model of hadronization. 
In the present paper, we assume that the particle content of the fireball at the
critical temperature $T_C$ can be found by considering 
a system of (non-interacting) hadrons in chemical equilibrium,
described by the grand canonical ensemble, which subsequently undergoes decay
processes. This is based on ideas presented in 
\cite{PBM1,PBM2,PBM3}, where it was shown that a huge number
of hadron ratios can be explained by a fit of only
two model parameters, temperature $T$ and baryochemical potential
$\mu_B$.

Clearly, the notion of non-interacting hadrons close to the
phase boundary is not suitable for the description of the
phase transition thermodynamics on a fundamental level, 
but should rather be regarded as an effective
prescription. The underlying assumption is that the
level-spacings and degeneracies of the QCD excitation spectrum
along with the energy scale provided by the temperature already
contain enough information to determine the relative
contributions of different hadron species and that details
of the interactions can either be absorbed into hadronic quasiparticle
properties such as temperature dependent masses and widths or
average out.

The calculation of hadronic abundancies in this framework
is a two step process: First, we have to specify the properties of
the medium (the QGP) at the transition in terms of the densities
of quarks, antiquarks and gluons (This is most conveniently done
by specifying the temperature $T_C$ and baryo- and strange chemical
potentials $\mu_B, \mu_S$). These parameters can be fit to the observed
hadron ratios (as done in \cite{PBM1,PBM2,PBM3}), calculated within
a dynamical model (as done in the present paper for SPS) or
inferred from experiment (as done in the present paper for RHIC).

Once the densities in the QGP have been specified, the hypothesis
of statistical hadronization is used to map those into the hadron ratios
observable in experiment. In detail, we use the following procedure to
calculate ratios for SPS:

Employing the grand canonical ensemble, 
we expect the density for each particle species $n_i$ to be given by
\begin{equation}
\label{E-GCE}
n_i = \frac{d_i}{2\pi^2}\int_0^\infty
\frac{p^2 dp}{\exp\{[E_i(p)-\mu_i]/T_C\}\pm 1}.
\end{equation}
Here, $d_i$ denotes the degeneracy factor of particle 
species $i$ (spin, isospin, particle / antiparticle), 
the +(-) sign is used for fermions (bosons) and 
$E_i(p)=\sqrt{m_i^2 + p^2}$. $m_i$ stands for the
particle's vacuum mass.
We use a value of 170 MeV for the critical temperature
$T_C$ as determined in lattice calculations
for two light and one heavy flavours \cite{Karsch-T_C}
and also used in the fireball evolution model.
The chemical potential $\mu_i$ takes care of conserved
baryon number $B_i$ and strangeness $S_i$ for each species:
\begin{equation}
\mu_i = \mu_B B_i - \mu_S S_i
\end{equation}
We neglect a (small) contribution $-\mu_{I_3} I_i^3$
coming from the isospin asymmetry in the colliding nuclei. 

The baryochemical potential $\mu_B$ is then fixed by
the requirement that the net number of baryons inside
the thermalized region is equal to the number of collision
participants $N_{part}$ if the total volume $V$ is known:
\begin{equation}
\label{E-Baryons}
V \sum_i n_i B_i = N_{part}.
\end{equation}
Similarly, strangeness conservation demands
\begin{equation}
\label{E-Strangeness}
V \sum_i n_i S_i = 0.
\end{equation}
Thus, the only parameter of the model is the fireball volume $V$.
At the phase transition, however, this volume can be determined
from the EoS of the QGP evaluated at $T=T_C$ using the entropy
density $s(T)$ as
\begin{equation}
\label{E-Volume}
V(T_C) = S_{tot}/s(T_C),
\end{equation}
if the total entropy content $S_{tot}$ of the fireball is known.
This quantity, however,  can be
obtained from measuring charged particle multiplicities $N^+$ and $N^-$ 
in suitable rapidity bins and
calculating
\begin{equation}
D_Q = \frac{N^+ - N^-}{N^+ + N^-} \label{D_Q}.
\end{equation}
The quantity $D_Q$ stands for the inverse of the specific entropy 
per net baryon $S/B$, and the product $D_Q(S/B)$ roughly measures the entropy 
per pion \cite{ENTROPY-BARYON}. For SPS collisions at 
160 AGeV, we find an entropy per net baryon $S/B = 26$ for
central collisons. 

At RHIC, hadron ratios are measured at midrapidity. It is known
that the distribution of baryon number across rapidity is
very inhomogeneous at collider energies, hence a prediction of
the quark and gluon densities within a globally averaged framework
as outlined for the SPS case using Eqs.~(\ref{E-Baryons}) and (\ref{E-Volume})
is bound to fail. For the sake of simplicity, we will not attempt to calculate
$\mu_B$ in a (more complicated) dynamical model  but rather infer it from
the experimentally accessible specific entropy end entropy, keeping in mind
that our results for RHIC do not test a dynamical model but
only the statistical hadronization hypothesis (in principle, however,
the baryon distribution in rapidity is calculable, e.g. \cite{BaryonDist}).
 
For RHIC 6\% central Au-Au collisions 
at 130 AGeV, the specific 
entropy $S/B = 220$ at midrapidity is substantially higher due to the
larger particle multiplicity 
and the smaller net baryon content in the central region. 
The transition volume is then estimated by using the EoS to
determine the volume at which the measured entropy in the midrapidity
slice correspond to the temperature $T_C$ if a backward extrapolation
of the expansion in time is made. Note, however, that the
transition volume is only relevant for excluded volume corrections and can be
neglected for a first order approximation  since
we are interested in hadron ratios and not total abundancies.

Thus, all ingredients entering Eq.~(\ref{E-GCE}) are determined
(without additional free parameters) and we can evaluate the expression for a suitable
choice of hadrons and resonances. 

We include all mesons and mesonic resonances up to masses of
1.5 GeV and all baryons and baryonic resonances up to masses
of 2 GeV.
This amounts to 30 (strange and nonstrange) mesonic states
and 36 (nonstrange to multistrange) baryonic states.
In order to compare to experimental results, we calculate
their decay into particles which are long-lived as compared
to the fireball, such as $\pi, K, \eta, N, \Lambda, \Sigma$ and
$\Omega$.

In order to account for interactions between particles, which at
small distances become repulsive, we assume a hard core radius
$R_{C}$ of 0.3 fm for all particles and resonances. The corresponding
excluded volume $V_{ex} = \sum_i N_i V_{ex}^i$,
with $V_i^{ex} = \frac{4\pi}{3} R_C^3$ for all species,  is subtracted 
from the volume obtained in Eq.~(\ref{E-Volume}), which in turn
affects the total number of produced particles. As the excluded
volume itself depends on the total number of particles, we iterate
the correction for a self-consistent result.
The hard core radius of 0.3 fm for protons is determined by comparison  with
p-p collisions \cite{HardCore}.  In the absence of such information
for the other mesons and baryons, we assume its universality.

All data on particles is taken from \cite{ParticleDataBook}.
For many higher-lying states, the properties as well as the
decay channels are poorly known. In these cases, we proceed
as follows: If  a quantity (e.g. masses and widths)
is given only within a certain range, the arithmetic
mean of this range is used in the model. Decay channels
which are reported to be 'seen' are assumed to receive
equal contributions from the branching ratio which is left
after all known channels have been accounted for.
Branching ratios less than 1\% have been neglected. Decay
chains (such as $a_2 \rightarrow \rho \pi \rightarrow \pi \pi\pi$)
have been followed through.  For resonances with large width,
we integrate Eq.~(\ref{E-GCE}) over the mass
range of the resonance using a Breit-Wigner distribution.

\section{Some remarks on thermodynamics}

As mentioned earlier, the notion of a non-interacting hadronic resonance
gas cannot be expected to describe the full thermodynamics of the
system close to the phase boundary. Therefore, one  cannot expect all
thermodynamical parameters to match smoothly coming from above and
below the phase transition. An
ensemble of free hadrons should therefore not be used to calculate thermodynamical
properties of the hot hadronic medium, as there is strong evidence
for re-scattering processes in the hadronic phase from
dilepton emission (see e.g. our result in \cite{MyDileptons}) and these
interactions certainly modify the thermodynamics.

This becomes apparent once we compare thermodynamic parameters directly.
Coming from above $T_C$, we find from the quasiparticle model of the QGP
\begin{displaymath}
p/T_C^4 = 0.3, \quad \epsilon/T_C^4 = 7.0 \quad \text{and} \quad s/T_C^3 = 7.3.
\end{displaymath}
Calculating the corresponding hadronic resonance gas values at $T_C$
(using the ensemble of particles and resonances described above), one
obtains
\begin{displaymath}
p/T_C^4 = 1.0, \quad \epsilon/T_C^4 = 6.3 \quad \text{and} \quad s/T_C^3 = 7.3.
\end{displaymath}
The crucial observation is that the entropy density matches smoothly
(one should keep in mind, however, that there are errors given by the 
extraction of $T_C$ from lattice data,
the extrapolation of lattice data to physical quark masses and
the poor knowledge of high-lying resonances).
This implies that we have not introduced a gross violation of the
number of available degrees of freedom (which is important, as the
goal of this study is to obtain particle numbers out of an
\emph{isentropic} expansion scenario with a crossover or second order
phase transition). If the entropy density did not match at $T_C$,
we would have introduced a first order phase transition which would certainly
lead to a breakdown of the simple, 'instantaneous' hadronization
scenario used here.

On the other hand, in the pressure and energy density we find discontinuities 
at $T_C$. The difference of the energy density in both phases
can be regarded as acceptable given the uncertainties, but in the 
case of the pressure the discrepancy is
large. It is, however, certainly conceiveable that attractive interactions
in the system lower the pressure. Thermodynamical
consistency then fixes the energy density via $\epsilon + p = sT$,
achieving continuity in both quantities.
A more fundamental approach in order to achieve continuity of all parameters across 
the phase transition, however, faces other difficulties:
\begin{itemize}
\item
Any interacting
model of hadrons will have to introduce additional coupling constants, which, at
least for the high-mass resonance states, are poorly known.
\item
Incorporating all \emph{known} interactions is of limited
help: If one considers the free ensemble at $T_C$, one
finds e.g. that the sum of all excited $\Delta$ states
is larger than the number of $\Delta(1232)$. This is
also true for the nucleon and nuclear resonances. 
Hence, resonance states are no
small contribution and cannot be neglected. If one is
interested in the number of pions resulting after resonance
decays, the role of heavy resonances is even further enhanced due
to very pion-rich decay channels of these heavy states.
\item 
Accepting the limited knowledge of heavy-resonance
properties, one might think of using some effective
\emph{ad hoc} prescription of restoring thermodynamical
consistency. However, there is no unique way of doing so,
therefore the model looses all predictive power.
\end{itemize} 

In contrast, statistical hadronization is a simple prescription
and completely determined by at most two parameters, which in 
the present context are given by the fireball evolution model.
Therefore, it appears useful to accept the limitations
of the present model for the benefit of a parameter-free
calculation.

\section{Results}

\begin{figure}[htb]
\begin{center}
\epsfig{file=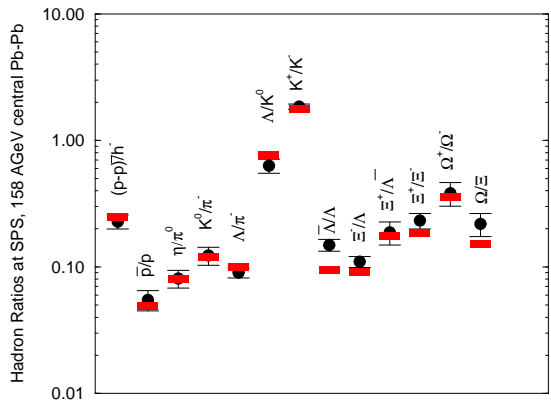, width=8cm}
\end{center}
\caption{\label{F-RatiosStd}Hadron ratios in the statistical hadronization
model (dashed bands) as compared to experimental results (filled circles) for SPS,
158 AGeV central Pb-Pb collisions.}
\end{figure}

The resulting hadron ratios are shown in Fig.~\ref{F-RatiosStd} for the
case of 158 AGeV central Pb-Pb collisions at SPS, compared with
the experimentally measured values \cite{R1,R2,R3,R4,R5,R6,R7,R8,R9,R10,R10a}. 
One observes
that the overall agreement with data is satisfactory with few exceptions, 
though not quite as good as in e.g. \cite{PBM2}
where a two parameter fit to the data was performed. Note that not
only the ratios of hadron yields agree to experiment in the present approach but also the
absolute numbers, as the baryochemical potential $\mu_B$ is explicitly linked
to the (known) number of participants.

The calculation yields a baryochemical potential $\mu_B = 265$ MeV and a strange chemical
potential $\mu_S = 59$ MeV. Note that these quantities depend on the number
of resonances included into the calculation; therefore, a direct comparison with
the values obtained in other types of models is not necessarily meaningful.

\begin{figure}[htb]
\begin{center}
\epsfig{file=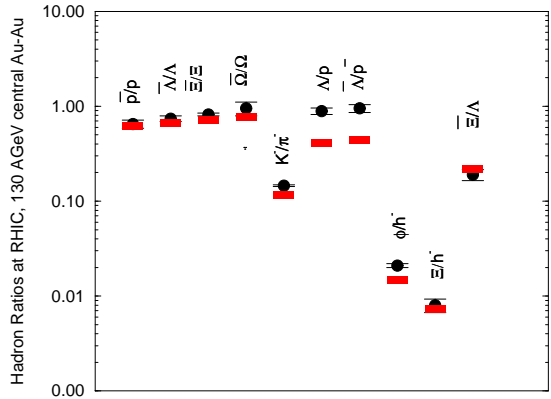, width=8cm}
\end{center}
\caption{\label{F-RatiosRHIC}Hadron ratios in the statistical hadronization
model at midrapidity (dashed bands) as compared to experimental results (filled circles) for RHIC,
130AGeV central Au-Au collisions at midrapidity.}
\end{figure}

For central collisions at RHIC at 130 AGeV beam energy   we present the results in Fig.~\ref{F-RatiosRHIC} and
compare to experimental data measured around midrapidity \cite{RHIC1,RHIC2,RHIC3,RHIC4,RHIC5,RHIC6}.
As the distribution of net baryon number in rapidity is very inhomogeneous at RHIC, the
naive application of Eq.~(\ref{E-Baryons}) fails, as we obtain a low
entropy per baryon, $S/B = 75$. At midrapidity, however, this ratio is closer to
$S/B = 220$. Using this value to calculate the ratios in a suitable interval around
midrapidity, we find much better agreement to the data (see Fig.~\ref{F-RatiosRHIC}).
Naturally, Eqs.~(\ref{E-Baryons}) and (\ref{E-Volume}) cannot be used in this case.
Instead, one has to consider the volume of the rapidity slice in
question and the number of baryons found inside this interval
(which can either be predicted by a model of measured in experiment).
Starting with this number and $S/B$, all ratios of produced hadrons within this
rapidity slice are then predicted. Clearly, a more detailed fireball evolution model
for the RHIC scenario is desirable to increase the predictive power.

\section{In-medium modifications}

In the last section, we have demonstrated that the idea of statistical hadronization
combined with our fireball evolution model is able to achieve good agreement 
in comparison with
data, provided that one uses the vacuum masses and widths of all resonances in
Eq.~(\ref{E-GCE}).  On the other hand, we have used the same fireball evolution
to calculate dilepton emission and here we found that a significant broadening of
the $\rho$ meson was responsible for the observed enhancement in the invariant
mass region below 700 MeV \cite{MyDileptons}.
The underlying thermal field theory calculations of the modifications
of the vector meson properties at finite temperature \cite{Omega,OldDileptons}
and density \cite{FiniteDensity} indicate not only a modified $\rho$
but also broadening and mass shift of the $\omega$ and broadening of the $\phi$
due to the interaction with the medium.

All these calculations (as any other perturbative expansion) become
unreliable in the vicinity of the phase transition, therefore these results
cannot strictly be taken over to the current calculations where we require
these quantities close to $T_C$. However, we may take them as a hint that
two possible modifications  of particle properties in a hot and dense medium may take
place:

\emph{Mass shifts} of  particles in the medium are commonly related to
the restoration of chiral symmetry. In e.g. \cite{BrownRho}, by investigating
scale invariance of an effective Lagrangian, the in-medium scaling laws
\begin{equation}
m_\rho^\ast/m_\rho \approx m_\omega^\ast/m_\omega \approx m_N^\ast/m_N \approx
\left(\langle \overline{q}q \rangle^\ast/\langle \overline{q}q \rangle \right)^{1/3}
\end{equation}
were established (Brown-Rho scaling). In this equation, asterisks denote
quantities at finite density, $m_\rho, m_\omega$ and $m_N$ the
masses of the $\rho$ and $\omega$ meson, $m_N$ is the nucleon mass
and $\langle \overline{q}q \rangle$ stands for the chiral condensate.

\emph{Decay widths} of particles are in general increased in a medium due to
the presence of new interaction channels. In \cite{Omega} e.g., it was shown that
the $\omega$ meson resonance experiences strong broadening in a hot environment due to
the presence of the scattering process $\omega \pi \rightarrow \pi \pi$.
Even in the absence of such effects, the decay width at finite temperature is
enhanced if the decay products are bosons due to the presence of
bosons of the same type in
the heat bath (Bose-enhancement). As pions are the most abundant species
in a thermal environment and most decays involve one or more pions, this
effect should influence almost all resonances in a hot medium.
The modification of decay widths is only relevant in Eq.~(\ref{E-GCE}) if
the in-medium width is a sizable fraction of the particle mass.
In this case, a large contribution to the particle yield comes from
masses lower than the peak mass in the Breit-Wigner distribution, which are
exponentially enhanced. This enhancement more than counterbalances
the suppression of contributions shifted into the higher mass
region. Therefore, an increase in the decay width of broad
resonances acts similarly as a mass reduction.

There is a third possibility of in-medium particle properties which has
no direct influence on the particle spectral function (and is
therefore not visible in the dilepton data):
The binding potential between the constituents of hadrons
could be partially screened by thermal fluctuations, leading to an
increase of the hadronic core radius.

Such \emph{increased radii} would lead to an enhanced excluded volume
correction. This effect hardly influences the results of
\cite{PBM1,PBM2,PBM3}, as the fireball volume is implicitly determined by
matching the fitted baryochemical potential to the number of participants,
but in the present approach we can expect to observe the influence of increased
hard core radii, as the volume is kept fixed.

\begin{figure}[!htb]
\begin{center}
\epsfig{file=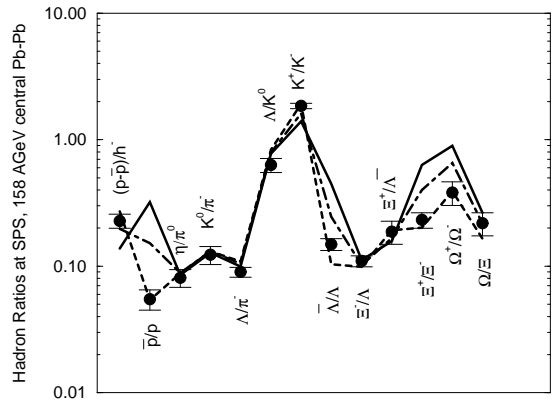, width=8cm}
\end{center}
\caption{\label{F-MVar}Hadron ratios in the statistical hadronization model, for
vacuum particle masses (dashed), assuming a reduction by 10\% (dash-dotted) and 20 \% (solid)
as compared to data (filled circles) for SPS, 158 AGeV central Pb-Pb collisions.}
\end{figure}

In a first step, we examine the effects of in-medium mass shifts in a qualitative
way by tentatively multiplying the vacuum masses of all hadrons with the exception of  the
pion by a constant $c$. The result is shown in Fig.~\ref{F-MVar}.

The result shows that even a moderate mass reduction of 10\% in the medium is not in line with
the observed hadron ratios. In particular, particle-antiparticle ratios are strongly affected.
In the case of $\overline{p}/p$, one might argue that the relevant inelastic
(annihilation) cross section is not small as compared to the elastic one and therefore
this ratio cannot really be fixed at $T_C$, but must be  adjusted dynamically in the
subsequent evolution. Indeed, it was shown in \cite{Rapp-ppbar} that this
is possible if one takes the statistical hadronization prediction as an initial
condition for rate equations. It is unclear if this is still possible for
different initial conditions, but even if this is the case, this is not an option
for the multistrange particle/antiparticle ratios.

The overall behaviour of the result can be qualitatively
understood as follows: The reduced nucleon mass implies a
lower value of the baryochemical potential in order to produce the observed number of participants,
this in turn affects single and double strange particles and implies changes in $\mu_S$ via
Eq.~(\ref{E-Strangeness}). Therefore, ratios of particles and antiparticles
with 2 or 3 non-strange valence quarks, 
such as $\overline{p}/p$
or $\overline{\Lambda}/\Lambda$ are most affected. Multistrange particle/antiparticle
pairs follow the trend, though in a way less pronounced.

\begin{figure}[!htb]
\begin{center}
\epsfig{file=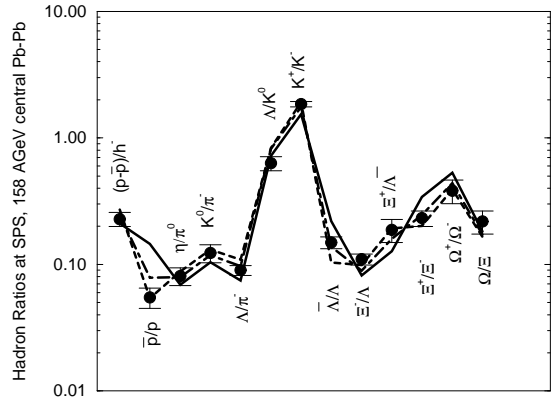, width=8cm}
\end{center}
\caption{\label{F-GammaVar}Hadron ratios in the statistical hadronization model, for
vacuum decay widths (dashed), assuming an increase by 20\% (dash-dotted) and 50 \% (solid)
as compared to data (filled circles) for SPS, 158 AGeV central Pb-Pb collisions.}
\end{figure}

In a second run, we investigate the effect of thermal broadening of resonances,
increasing all widths by a constant multiplicative factor $c_\Gamma$. The resulting
hadron ratios are shown in Fig.~\ref{F-GammaVar}.

One observes the same qualitative behaviour as for a mass reduction, as we have argued before.
The effects of the increased width are, however, less dramatic. An increase by 20\% in the
width of all resonances is still in line with all data except $\overline{p}/p$ and even
an increase of 50\%  is still acceptable for most of the ratios. This is reassuring, as there
is almost certainly thermal broadening of resonances in a hot medium.

As the behaviour for both broadening of resonances and mass reduction (which are
both in line with the dilepton data) is qualitatively the same for the hadron ratios,
the effects of reduced masses cannot be compensated by introducing additional
broadening. So if any of the effects fails in the description of the data, a combination
of both will also fail. 

\begin{figure}[!htb]
\begin{center}
\epsfig{file=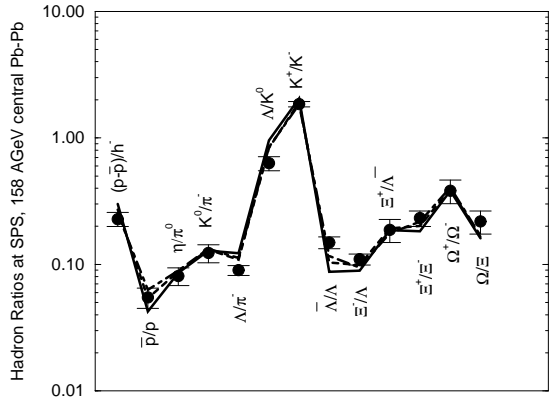, width=8cm}
\end{center}
\caption{\label{F-Rvar}Hadron ratios in the statistical hadronization model, for
the standard choice of $R_C = 0.3$ fm (dashed), assuming $R_C = 0.225$ fm (dash-dotted) and
$R_C = 0.375$ fm (solid)
as compared to data (filled circles) for SPS, 158 AGeV central Pb-Pb collisions.}
\end{figure}

In the third run, we explore the effect of different choices for the core radius $R_C$
on the ratios. In order to investigate the sensitivity of our results to
the initial choice of $R_C$, we do not only consider thermally increased radii
but also a reduced initial choice. In Fig.~\ref{F-Rvar}, we show the model predictions
for a reduction of $R_C$ by 25\% and for an increase of the same amount.

We observe that the overall sensitivity of the resulting hadron ratios
to the hadronic  core radius is rather weak. On the other hand, an enhanced
excluded volume
correction (dotted line in Fig.~\ref{F-Rvar}) acts in a rather peculiar way:
In order to arrive at the same number of participants, the baryochemical
potential has to \emph{increase}. This effect is opposite to the behaviour 
for in-medium mass reductions or increase of decay width. Therefore,
one might expect that the net effect of both thermal broadening of resonances \emph{and}
increased core radius due to a thermally screened binding potential
is moderate and partial compensation may occur.

\begin{figure}[!htb]
\begin{center}
\epsfig{file=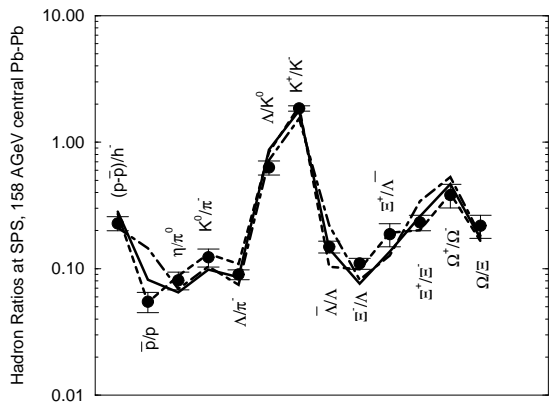, width=8cm}
\end{center}
\caption{\label{F-Radius}Hadron ratios in the statistical hadronization model assuming
all decay widths increased by 50\%, for
the standard choice of $R_C = 0.3$ fm (dash-dotted) and $R_C = 0.45$ fm (solid)
as compared to data (filled circles) for SPS, 158 AGeV central Pb-Pb collisions.}
\end{figure}

In order to test this conjecture, we study a scenario in which
the width of resonances has been increased by 50\% and simultaneously the
core radius has been set to 0.45 fm instead of 0.3 fm. The result is
shown in Fig.~\ref{F-Radius}.

One observes that indeed the expected compensation occurs and the model
prediction approaches the data points.  No effort has been made to
obtain a best description of the data using $c_\Gamma$ and $R_{hc}$ as
fit parameters.

\section{Conclusions}

Assuming a thermalized system and statistical hadronization at the phase boundary
with subsequent resonance decays,
we have demonstrated that our recently proposed fireball evolution
scenario \cite{MyDileptons} leads to a reasonable description of the data
on hadron ratios, if the known vacuum properties of particles
and resonances are used. This provides a valuable consistency
check of three pieces of information: 1) The entropy density at
$T_C= 170$ MeV can be calculated using the hadronic ensemble.
The entropy density of the QGP has to agree with that number
in order to produce a second order transition or a crossover,
as assumed in the fireball evolution scenario.
2) The quasiparticle model predicts the entropy density
at $T_C$, but without specifying an absolute value for
$T_C$.  3) The essential information of the absolute
value of $T_C$ is obtained in lattice calculations
\cite{Karsch-T_C}, and using this value, we find a
consistent scenario. In that sense, 
hadron ratios provide information on the EoS
of the quark-gluon plasma.

On the other hand, in-medium modifications of hadron properties are mandatory
if one tries to explain the dilepton invariant mass spectrum
measured by the CERES collaboration \cite{CERES}. Specifically, the $\rho$ channel
requires strong broadening.
Thermal effective field theory calculations indicate the presence of such
effects for other particles as well \cite{OldDileptons, FiniteDensity}.

In a schematic investigation, we have demonstrated that such strong
broadening of resonances or mass reductions are not in line with
the measured data, and no combination of these two effects can be.
Such changes in particle properties are also
inconsistent with the phase transition scenario
used here, as they destroy the match of the entropy density and require in general
a different choice of $T_C$. This is also apparent from the
fact that the \emph{absolute} numbers of particle production
do not agree with measurements. 

However, if one also consideres the effect of a screened binding
potential in a thermal environment, one can make the assumption that
the core radius
of hadrons grows. This effect can then compensate the
effects of both increased decay widths and mass reductions
to some degree, as demonstrated in Fig.~\ref{F-Radius}. 
This result is reassuring, as it allows to reconcile the
in-medium modifications observed in the dilepton data with
the statistical model description of measured hadron ratios in a
consistent model framework.

\section*{Acknowledgements}

I would like to thank W.~Weise, A.~Polleri, R.~A.~Schneider, P.~Braun-Munzinger and J.~Stachel
for interesting discussions and helpful comments.

\end{document}